\documentclass{article}

\usepackage{amssymb}
\usepackage{amsfonts}
\usepackage{amsmath}
\usepackage{graphicx}
\usepackage{hyperref}
\hypersetup{backref,
            pagebackref=true,
            ps2pdf,
            bookmarks=true,
            bookmarksnumbered=true,
            pdfauthor=Edward G.J. Lee,
            colorlinks=true,
            linkcolor=blue}

\title{Excitation Spectrum at the Yang-Lee Edge Singularity
of the 2D Ising model}

\author{\textrm{TOMASZ WYDRO}\\%
\textit{Laboratoire de Physique Mol\'{e}culaire et des Collisions, Universit\'{e} de Metz,} \\
\textit{1 bvd Arago, 57078 Metz,France}\\
wydro@sciences.univ-metz.fr
\\ \\
\textrm{JOHN F. McCABE}\\
\textit{412 Morris Ave., \# 34, Summit, NJ 07901,USA}\\
jfmccabe2@earthlink.net
}
\begin{document}

 \maketitle

\begin{abstract}
This paper studies the Yang-Lee edge singularity of 2-dimensional
$(2D)$ Ising model through a quantum spin chain. In particular,
finite-size scaling measurements on the quantum spin chain are used
to determine the low-lying excitation spectrum and central charge at
the Yang-Lee edge singularity. The measured values are consistent
with predictions for the $(A_{4}, A_{1})$ minimal conformal field
theory.
\end{abstract}

\begin{center}
\bigskip
\end{center}

In 1978, Fisher\cite{Fisher1} argued forcefully that Yang-Lee edge
singularities\cite{YL1,YL2} are similar to ordinary critical points.
Later, Cardy\cite{IsingYLS} provided another strong argument that
the Yang-Lee edge singularity of the 2D Ising model is identified
with the $(A_{4}, A_{1})$ minimal conformal field theory
(CFT)\cite{BPZ,Friedan} of the ADE classification.\cite{ADE1}
Cardy's identification enables a number of CFT predictions about the
Yang-Lee edge singularity of the 2D Ising model.

Some of these CFT predictions have been confirmed. Using transfer
matrix \mbox{methods,} Itzykson et al measured the central charge
and exponent $\nu$ at the Yang-Lee edge singularity of the 2D Ising
model.\cite{Zuber} Using short quantum spin chains,\cite{Fradk78}
Uzelac et al also measured the exponent $\nu$ at the Yang-Lee edge
singularity of the 2D Ising model.\cite{Uzelac81} Both measurements
are in agreement with predictions derived from Cardy's
identification.

This article provides other measurements that can be predicted based
on Cardy's identification. The measurements provide the low-lying
excitation spectrum at this Yang-Lee edge singularity of the 2D
Ising model. This article also compares the measured low-lying
excitation spectrum with predictions based on Cardy's identification
of the $(A_{4}, A_{1})$ minimal CFT with this Yang-Lee edge
singularity of the 2D Ising model.\cite{IsingYLS,Zuber,Uzelac81}

In this article, measurements\cite{FiniteScall} are based on
finite-size scaling in quantum spin chains. For the 2D Ising model
in an imaginary external magnetic field, the associated $N$-site
quantum spin chain has a Hamiltonian, $H_{Ising}$, given
by:\mbox{\cite{Gehlen87}}
\begin{equation}
H_{Ising}=-\sum_{n=1}^{N}\{t\sigma_{z}(n)\sigma_{z}(n+1)+iB\sigma_{z}(n)+\sigma_{x}(n)\}\text{
.} \label{HIsing}
\end{equation}
In Eq. (\ref{HIsing}), $\sigma_{x}(n)$ and $\sigma_{z}(n)$ are 2x2
Pauli spin matrices at site $n$, the parameter "$t$" is a positive
coupling for ferromagnetic spin-spin interactions, and $iB$ is a
purely imaginary external magnetic field.  In Eq. (\ref{HIsing}),
the last term results from single inter-row spin flips in the
associated 2D transfer matrix.\mbox{\cite{W-MC1}}

In this article, the phenomenological renormalization group (PRG) is
used to determine critical values of the purely imaginary magnetic
field, $iB_{YL}(N)$, for various chain lengths, N. For such purely
imaginary magnetic fields, the PRG equation
becomes:\cite{Derrida,Zuber}
\begin{equation}
[N -1]m(B_{YL}(N), N-1)= [N]m(B_{YL}(N), N)\text{ .} \label{PRG}
\end{equation}
In Eq. (\ref{PRG}), $m(B, N) = \left[E_{1}(B, N) - E_{0}(B,
N)\right]$ where $E_{0}(B, N)$ and $E_{1}(B,N)$ are the energies of
the ground state "0" and first excited state "1" of the quantum spin
chain having length N. At the $B_{YL}(N)$'s, the Ising quantum spin
chain exhibits finite-size scaling behavior associated with the
Yang-Lee edge singularity of the 2D Ising spin model. In particular,
if the $B_{YL}(N)$'s converge to a nonzero value as
$N\rightarrow\infty$, that value corresponds to the critical point
for the Yang-Lee edge singularity of the 2D Ising spin model.

At the $B_{YL}(N)$'s, excitation energies and other physical
quantities should scale. CFT predicts the scaling behavior of such
physical quantities at the $B_{YL}(N)$'s.

First, CFT predicts how excitation energies will scale with the
length, N, of the quantum spin chain. For an excited energy
eigenstate "i" of the quantum spin chain, the CFT prediction is that
the excitation energy, $E_{i}(N)-E_{0}(N)$, will scale
as:\cite{Cardy1-84}
\begin{equation}
E_{i}(N)-E_{0}(N)=\zeta 2\pi
\frac{\Delta_{i}+\bar{\Delta}_{i}-(\Delta+\bar{\Delta})}{N}\text{ .}
\label{gaps scaling}
\end{equation}
In Eq. (\ref{gaps scaling}), $\Delta_{i}$ and $\bar{\Delta}_{i}$ are
left and right conformal dimensions of the field "i" in the
associated CFT, and $\Delta$ and $\bar{\Delta}$ are the conformal
dimensions of the primary field of lowest "negative" scaling
dimension in the CFT. Such negative dimension fields occur in
various non-unitary CFTs. In Eq. (\ref{gaps scaling}), the constant
$\zeta$ is non-universal and depends, e.g., on the normalization of
the Hamiltonian of the quantum spin chain.

Second, CFT predicts how the ground state energy, {$E_{0}(N)$}, will
scale with the length, N, of the quantum spin chain. In particular,
the ground state energy scales as:\cite{Cardy84,Reinicke}
\begin{equation}
E_{0}(N)= AN + B -\zeta\pi {C_{eff}}/(6N)+\ldots
 \label{ground state scaling}
\end{equation}
While the constants A and B are non-universal, ${C_{eff}}$ and the
exponents of the finite-size corrections in 1/N of the last two
terms are universal numbers predicted by CFT.  In particular,
${C_{eff}}$, is the effective central charge of the CFT.
${C_{eff}}$ is defined by ${C_{eff}} = C - (\Delta + \bar\Delta)$
where "$C$" is the normal central charge of the CFT.\cite{Zuber}

For the minimal CFTs, the ADE classification provides modular
invariant partition functions\cite{ADE1,RochaCarida} from which the
low-lying excitation spectrum and central charge are easily
extracted. For the $(A_{4}, A_{1})$ minimal CFT, Table
\ref{tab:Spectra} shows the low-lying excitation energies and
associated
degeneracies of the spectrum obtained from the associated
modular invariant partition function of the ADE classification.
\begin{table}[h]
\centering
\begin{tabular} {|c||c|c|c|c|c|c|c|c|c|}
\hline \multicolumn{1}{|c||}{CFT}&
\multicolumn{6}{c|}{$(A_{4},A_{1})$}\\
\hline
Normalized Energies &0&1&  2.5& 5.0& 6.0&   7.5\\
\hline
Degeneracy  &1&1&    2&    3&    2&     4\\
\hline
\end{tabular}
\caption{Low-lying excitation spectrum of $(A_{4}, A_{1})$ minimal CFT \label{tab:Spectra}}
\end{table}
Rather than absolute excitation energies, Table \ref{tab:Spectra}
lists normalized excitation energies, i.e., ratios. For a state "i",
the associated normalized excitation energy is defined as the ratio
is the actual excitation energy of the state "i" over the actual
excitation energy of the lowest excited state "1". Here, absolute
excited energies are measured with respect to the ground state
energy. The ratios of Table \ref{tab:Spectra} have the advantage of
not depending on non-universal constants such as $\zeta$. For the
$(A_{4}, A_{1})$ minimal CFT, the form of the modular invariant
partition function also implies that the effective central charge,
$C_{eff}$, is equal to 2/5.\cite{Zuber} Below, finite-size scaling
measurements on the Ising quantum spin chain are used to determine
the low-lying excitation spectrum and effective central charge at
the Yang-Lee edge singularity.

The measurements of the critical magnetic fields, $B_{YL}(N)$, were
obtained by numerically solving PRG eq. (\ref{PRG}) for Ising
quantum chains of different lengths.  For the numerical solutions,
the state energies were obtained by applying the Lanczos algorithm
to $H_{Ising}$ of eq. (\ref{HIsing}). Table \ref{tab: HC} shows the
values for the critical fields, i.e., $B_{YL}(N)$'s, ground state
energies, and lowest excitation energies, i.e., Gap(N)'s, as
obtained via the Lanczos algorithm. Table \ref{tab: HC} lists
measurements of these physical quantities for Ising quantum spin
chains in which the coupling, t, had the value of 0.1. The
measurements of Table \ref{tab: HC} also show that NxGap(N) scales
to a constant as $N\rightarrow\infty$ as expected for the PRG.

\begin{table}[h]
{
\centering
\begin{tabular} {|c|c|c|c|c|}
\hline
 Number& & & & \\
of Sites &  $B_{YL}(N)$ & Energy of ground state & Gap(N) & $N \times Gap(N)$\\
\hline

3 & .2459180i  & -2.8811043 & .8103423  & 2.4310 \\
4 & .2384127i  & -3.8028211 & .6629112  & 2.6516 \\
5 & .2352339i  & -4.7341982 & .5613016  & 2.8065 \\
6 & .2337637i  & -5.6688215 & .4858628  & 2.9152 \\
7 & .2330279i  & -6.6048003 & .4275400  & 2.9928 \\
8 & .2326347i  & -7.5414746 & .3811698  & 3.0494 \\
9 & .2324118i  & -8.4785910 & .3435105  & 3.0916 \\
10& .2322793i  & -9.4160213 & .3123765  & 3.1237 \\
11& .2321972i  & -10.353696 & .286250   & 3.1488 \\
12& .2321442i  & -11.291568 & .264041   & 3.1685 \\
 ...   & ...   & ...        & ...       & ...    \\
$\infty$&.23193i&$-\infty$  & 0.0       & 3.2840 \\
\hline
\end{tabular}}
\caption{PRG results for $B_{YL}(N)$, Ground state energy, Gap, and
NxGap as a function of the number of sites, N \label{tab: HC}}
\end{table}

Based on the finite-size scaling measurements of Table \ref{tab:
HC}, we first evaluated the effective central charge, $C_{eff}$, at
the Yang-Lee edge singularity of the 2D Ising model to test the
numerical methods. In particular, ground state energies for adjacent
triplets of chain lengths [(N-1), N, (N+1)] and an estimate,
$\zeta(N)$, for the non-universal constant, $\zeta$, were used to solve
Eq. (\ref{ground state scaling}) for estimates, i.e.,
$C_{eff}(N)$'s, to the effective central charge, $C_{eff}$. In each
evaluation, the Gap(N) was used to estimate the non-universal
constant $\zeta(N)$ of Eq. (\ref{gaps scaling}) at chain length "N".
The resulting estimates, i.e., the $C_{eff}(N)$'s, are plotted in
Figure 1. The $C_{eff}(N)$'s have a "1/N" dependence due to the
higher order "1/N" corrections to eqs. (\ref{gaps scaling}) and
(\ref{ground state scaling}).
\begin{figure}[h,t,p]
\centering
    \includegraphics{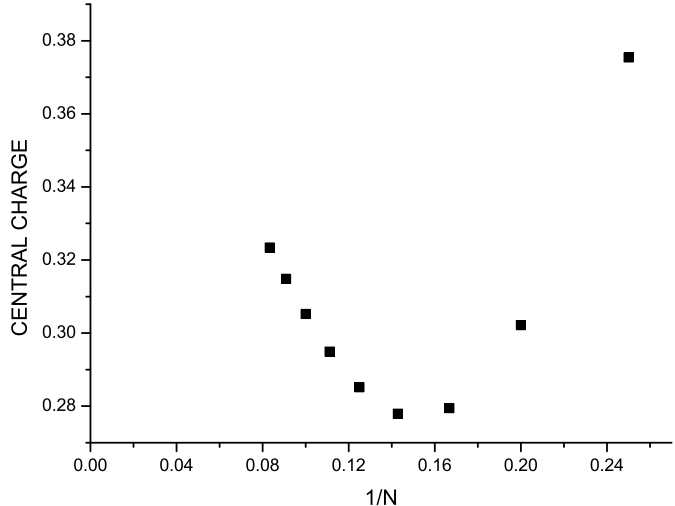}
  \caption{The $C_{eff}(N)$ estimates for the effective central
  charge, $C_{eff}$, as a function of 1/N.}\label{fig:Ceff}
\end{figure}

From a visual inspection of Figure 1, one sees that the estimates
for the effective central charge have a crossover behavior in $1/N$.
After the crossover, the $C_{eff}(N)$'s converge smoothly towards a
limit, i.e., $C_{eff}$, as $1/N \rightarrow 0$. The effective
central charge may be obtained by extrapolating the estimates
towards 1/N = 0.  A visual inspection of Figure 1 also shows that
the $C_{eff}(N)$'s behave as $C_{eff} + \alpha/N$ as $1/N
\rightarrow 0$.  Based on this form, one easily extracts from the
measured values that $C_{eff}$ is approximately equal to 0.41. This
measured value for $C_{eff}$ agrees well with the value of 2/5 that
the $(A_{4}, A_{1})$ minimal CFT predicts for $C_{eff}$.

The critical magnetic field values of Table \ref{tab: HC}, i.e., the
$B_{YL}$'s, were also used to find the low-lying excitation spectra
of Ising quantum spin chains of various lengths. While the Lanczos
algorithm can provide the low-lying excitation spectra, it is
inconvenient for determining the complete low-lying excitation
spectrum. In particular, the Lanczos algorithm requires one to
select a starting vector and then, produce a Krylov space over the
starting vector.  For simple choices of the starting vector, we
found that the Lanczos algorithm spreads the low-lying excitations
over several Krylov spaces. Thus, an evaluation of the complete
low-lying excitation spectrum via the Lanczos algorithm would
require finding a first Krylov space and then, finding other Krylov
spaces orthogonal to the first space.  To avoid such numerical
complications, we used the commercial linear algebra package of
Maple$_{TM}$ 9.5 to generate complete excitation spectra.

The Maple$_{TM}$ package provided excitation spectra for Ising
quantum spin chains with 6 - 12 sites. The measured low-lying parts
of said spectra including both energies and degeneracies are
provided in Table \ref{tab: mp}.  There, the excitation energies are
again normalized by division by the lowest excitation energy of the
Ising quantum spin chain.  As already described, such a
normalization removes any dependence on the non-universal constant
$\zeta$.

\begin{table}[t b p]
\centering
\begin{tabular} {|c||c|c|c|c|c|c|c|}
\hline
State /[Degeneracy] &  6    &  7     &  8   &   9    &     10   &    11   &  12  \\
 \hline
A / [2] & 2.68432  & 2.64386 & 2.61415 & 2.59207 & 2.57540 & 2.56260 & 2.55253 \\
\hline
B / [1] & 4.18193  & 4.27896 & 4.36713 & 4.44474 & 4.51197 & 4.56977 & 4.61912 \\
\hline
C / [2] & 4.51738  & 4.63236 & 4.70368 & 4.75182 & 4.78652 & 4.81281 & 4.83329 \\
\hline
D / [2] & 5.85889  & 5.89208 & 5.91240 & 5.92644 & 5.93703 & 5.94544 & 5.95210 \\
\hline
E / [2] &    --    & 5.68559 & 6.03104 & 6.27270 & 6.45018 & 6.58573 & 6.69223 \\
\hline
F / [2] &    --    & 6.24252 & 6.35798 & 6.46344 & 6.55966 & 6.64694 & 6.72535 \\
\hline
\end{tabular}
\caption{The table shows normalized low-lying excitation energies of
states A - F for Ising quantum spin chains with 6 to 12 sites.
Degeneracies of the states are as listed in the left column. \label{tab: mp}}
\end{table}

From the measured spectra of Table \ref{tab: mp}, one can determine
the form of the low-lying excitation spectrum in the limit where
$1/N \rightarrow 0$. The determination requires first, identifying
classes of eigenstates that correspond for different chain lengths
and then, determining the scaling behavior for each of the
identified classes as, $1/N\rightarrow 0 $. The identification of
corresponding eigenstates for different values of N is achieved with
the aid of some simple rules. The first rule is that the energies of
corresponding states vary monotonically and smoothly with N provided
that the initial value of N is sufficiently large. The second rule
is that the correspondences between states must account for
degeneracies. In particular, while new eigenstates appear as N
increases, old eigenstates do not disappear.\footnote{A few
"special" low-lying excited eigenstates have excitation energies
whose magnitudes grow rapidly with N. The"special" eigenstates are
not part of the low-energy excitation spectrum as $N\rightarrow
\infty $.  In fact, the number eigenstates with energies of smaller
magnitude than those of the "special" eigenstates grows with N. Such
"special" states were also seen at the Yang-Lee edge singularity of
the 3-state Potts quantum spin chain where they were not part of the
spectrum in the thermodynamic limit.\cite{W-MC2} The "special"
excited eigenstates are not discussed further.} Thus, in each class,
degeneracies of eigenstates either stay constant or increase with N.

Application of the above rules leads the classification of low-lying
excited eigenstates of Table \ref {tab: mp}.  The classification
includes classes A - F of state types.\footnote{Here, the lowest
excited state has been ignored. The lowest excited eigenstate was
however, observed to be non-degenerate as predicted for the $(A_{4},
A_{1})$ minimal CFT.  The lowest lying excited eigenstate
automatically has an energy of 1 in the convention of Table \ref
{tab:Spectra}. } Figures 2 - 5 plot the normalized excitation
energies for the states of classes A - F as a function of the
inverse of the length of the Ising quantum spin chain.

\begin{figure}[h t p]
    \includegraphics{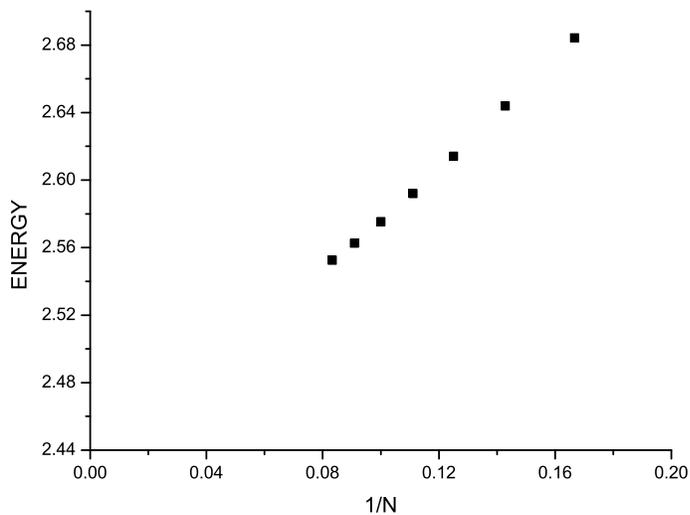}%
    \caption{The measured energies of type A states are plotted in 1/N.}%
    \label{fig:A}
\end{figure}

\begin{figure}[h t p]
    \includegraphics{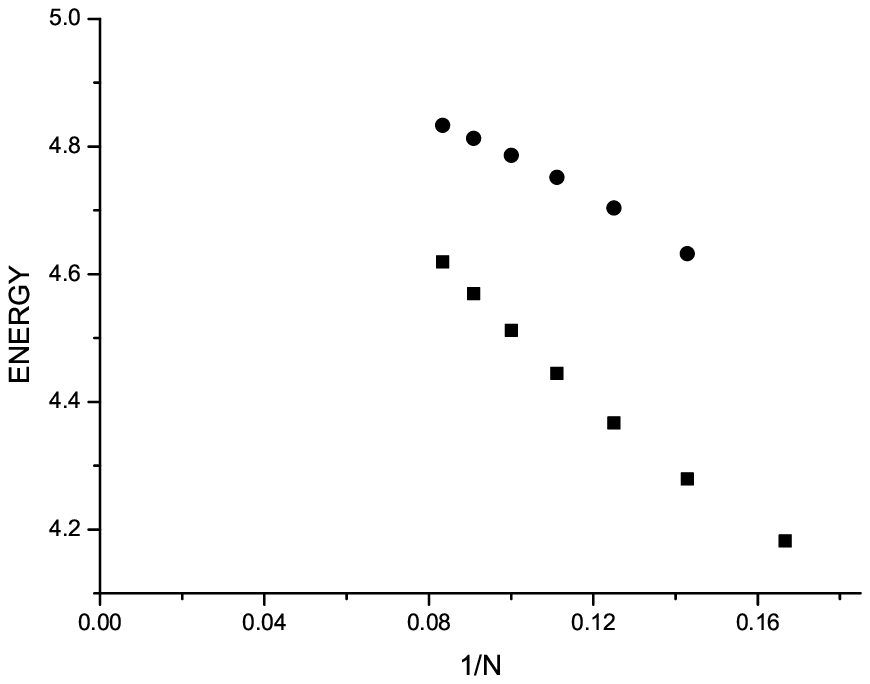}%
    \caption{The measured energies of type B states (squares) and of type C states (circles)  are plotted in 1/N.}%
    \label{fig:BandC}
\end{figure}

\begin{figure}[h t p]
    \includegraphics{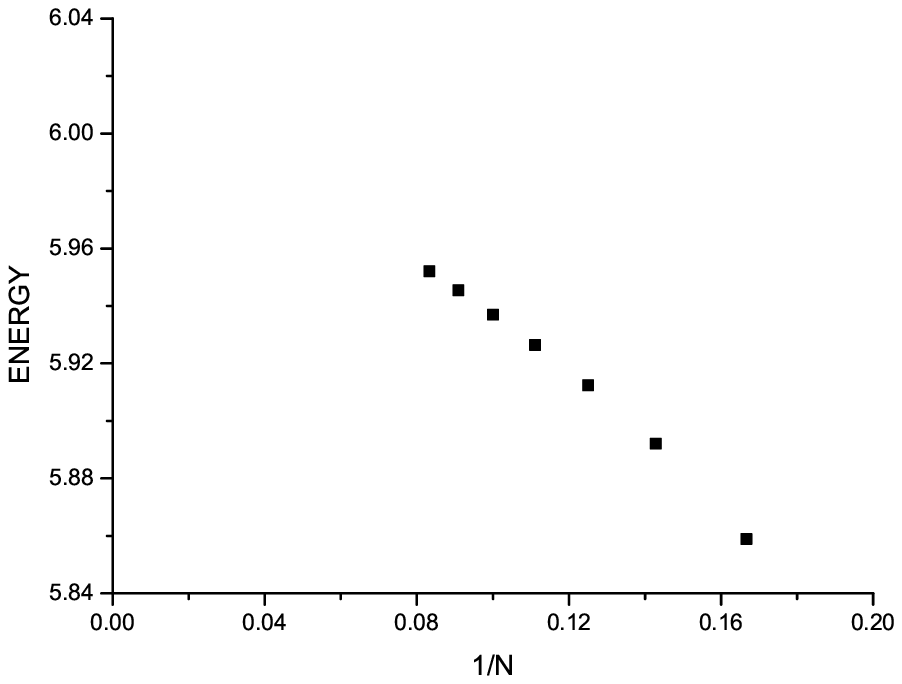}%
    \caption{The measured energies of type D states are plotted in 1/N.}%
    \label{fig:D}
\end{figure}

\begin{figure}[h t p]
    \includegraphics{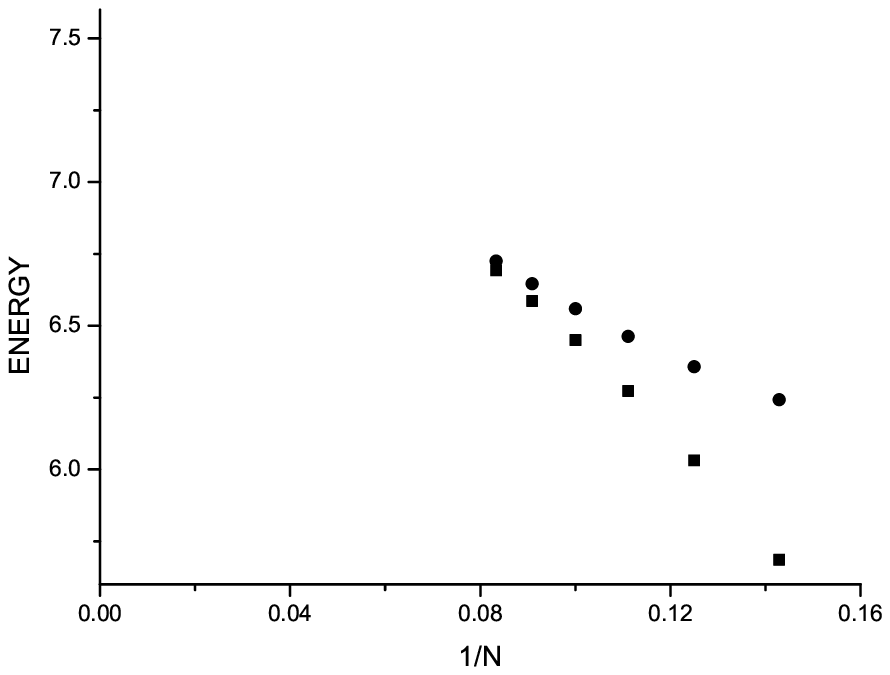}%
    \caption{The measured energies of type E states (squares) and of type F states (circles) are plotted in 1/N.}%
    \label{fig:EandF}
\end{figure}

A visual inspection of Figures 2 - 5 readily shows that the excited
states A, B, C, D, E, and F combine into the four distinct sets A, B
$\&$ C, D, and E $\&$ F.  Within each distinct set, the excited
eigenstates have energies that approach the same value as $1/N
\rightarrow 0$.

A visual inspection of Figures 2 - 5 shows that the normalized
energies of the states of the sets A, B $\&$ C, D, and E $\&$ F
approach about 2.45, 5.0, 6.03, and 7.6, respectively, as $1/N
\rightarrow 0$.  A BST analysis\cite{BSTA} was performed to obtain
more reliable values for the energies of these states in the limit
where $1/N \rightarrow 0$. The BST analysis indicated that the
normalized energies of the states of type A, B, C, D, E, and F scale
to 2.4995(5), 5.005(1), 5.003(3), 5.99(1), 7.54(8), and 7.60(7),
respectively, in this limit.\footnote{The BST analysis also provided
evidence that finite-size scaling corrections to the normalized
energies of these sets of states scale as $1/N^b$ where the b's are
in [1, 2] as $1/N \rightarrow 0$.}  These finite-size scaling
measurements of the low-lying excitation energies agree well with
those of the $(A_4, A_1)$ CFT as shown in Table \ref{tab:Spectra}.

Also, an inspection of Table \ref{tab: mp} shows that the distinct
state sets A, B $\&$ C, D, and E $\&$ F, of different limiting
excitation energies have 2, 3, 2, and 4 states, respectively. Thus,
the PRG measurements of the normalized excitation spectra also
provide values for the degeneracies that agree well with those of
the $(A_4, A_1)$ CFT of Table \ref{tab:Spectra}.

In conclusion, our finite-size scaling measurements on the Ising
quantum spin chain at the Yang-Lee edge singularity have produced a
low-lying excitation spectrum that is in very good agreement with
that predicted from the $(A_4, A_1)$ minimal CFT.  These results
further confirm Cardy's identification of the Yang-Lee edge
singularity of the 2D Ising model with the $(A_{4},A_{1})$ minimal
CFT.

\end{document}